\documentclass[12pt]{article}
\usepackage{amssymb}
\usepackage{graphicx}

\begin{document}

\title{Small World Graphs by the iterated ``My Friends are Your Friends'' Principle}
\author{Ph.Blanchard, A.Ruschhaupt and T.Krueger \\
\textit{University of Bielefeld}\\
\textit{Faculty of Physics, Germany}}
\date{ }
\maketitle

\begin{abstract}
We study graphs obtained by successive creation and destruction of edges
into small neighborhoods of the vertices. Starting with a circle graph of
large diameter we obtain small world graphs with logarithmic diameter, high
clustering coefficients and a fat tail distribution for the degree. Only
local edge formation processes are involved and no preferential attachment
was used. Furthermore we found an interesting phase transition with respect
to the initial conditions.
\end{abstract}

%
%

\section{Introduction}

In 1998 Watts and Strogatz introduced in a seminal paper Small World
Networks and provided a simple model of such graphs \cite{1,5}. The key
property of Small World Graphs is the simultaneous presence of small
diameter (at most logarithmic in the vertex size) and high clustering
coefficient. Extensive investigations over the last decade of real networks
like the WorldWideWeb, citation networks, friendship networks and many
others have shown that these graphs are really of Small World type. But
there is a second remarkable property shared by most real networks namely
that they have a power law like fat tail distribution (see \cite{4} for an
excellent recent survey about the whole subject).

There are many models which produce graphs with small diameter and power law
distribution. If additionally local search rules are incorporated one easily
obtains high clustering as well. But up to now it was common believe that
for the small diameter property a certain amount of essentially independent
random edge formations like in classical random graphs where necessary. To
obtain a scale free distribution for the degree the basic conviction was
that the random edge formation should be biased according to the ``the
richer you are the richer you get- principle'' \cite{3} where richness is
just measured by the degree of a vertex. In the present paper we will
present a model which shows that even by entirely local edge formation rules
without preferential choices graphs are obtainable which admit all the three
above mentioned properties. The results are entirely based on numerical
simulations since an analytic treatment of the model seems to be rather
difficult. Nevertheless we hope to present in a forthcoming paper a more
theoretical study of the main properties.

%
%

\section{The iterated ``My Friends are Your Friends'' Principle}

Looking at the edge formation process in social networks there are essential
two ways to get a new contact. First by the random event of meeting somebody
for instance in a train or airplane. Humans refer to this kind of relation
creation often as fate. Second there are contacts which where created by the
local rule ``let me introduce you to one of my friends'', certainly a very
common process in real life. In the language of graphs this translates to
the formation of an edge between two vertices say $x$ and $y$ which had
distance 2 with respect to the underlying graph metric and we refer to this

kind of edge formation as the ``My Friends are your Friends'' Principle
(FFP). Graphs which where partially build by using the FF-Principle are
already described and analyzed in \cite{2}. In the following we want to
describe a model based on the iteration of this principle - the IFF-model. A
typical example of an essential FF-graph (although a directed one and with
increasing number of vertices) is the citation network.

%
%

\section{The model description and results}

In detail, we want to discuss the following model of a time dependent random
graph space $\mathcal{G}_{t}\left( N,t_{0}\right) $ with $N$ vertices and an
explicit integer parameter $t_{0}$ tuning the total number of edges. We
start with a circular chain of $N$ vertices with the usual nearest neighbor
edges, i.e. vertex $1$ is connected to vertex $N$ and $2$, vertex $2$ is
connected to vertex $1$ and $3$ and so on. We call this configuration a
circle and refer to the corresponding edges as fixed base edges. The random
graph space $\mathcal{G}_{t}\left( N,t_{0}\right) $ will be defined through
the following algorithm: In each of the first $t_{0}$ time-steps, a vertex
say $x$ is chosen randomly. From the set of neighbors $N_{1}\left(
x,t\right) $ of this vertex (at the given time $t$) one element say $y$ is
chosen randomly and then randomly from $N_{1}\left( y,t\right) \setminus
\left\{ x\right\} $ -- the neighbor set of $y$ at time $t$ with the
exclusion of $x$ -- another element say $z$. If there was no edge between $z$
and $x$ a new one is created between the two vertices otherwise nothing
happens. In the first case we say that $x$ has created an FF-edge to $z-$
respectively $z$ was chosen by $x$. After repeating $t_{0}$ times the above
procedure one continues for $t>t_{0}$ the same way with the additional rule
that whenever a vertex say $x$ has created a new FF-edge randomly an FF-edge
containing $x$ is deleted. Note that the total number of edges stays
therefore constant for $t>t_{0}.$

\subsection{The diameter}

\begin{figure}[tbp]
\begin{center}
\includegraphics {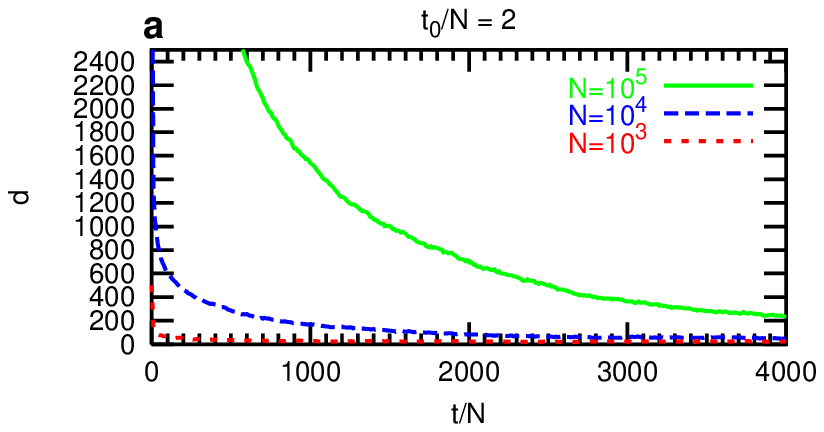}
\par
\includegraphics {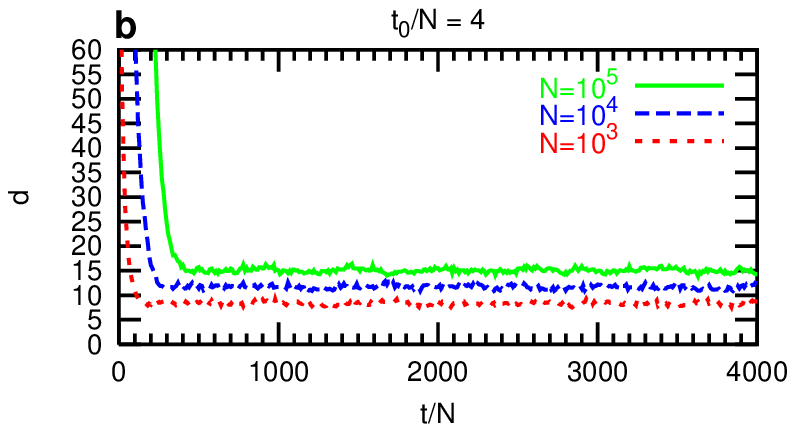}
\par
\includegraphics {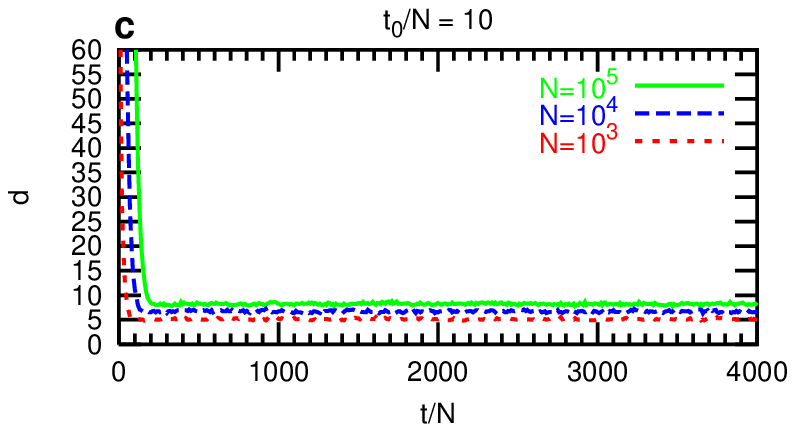} \vspace{-1cm}
\end{center}
\bf\caption{\rm Diameter of $\mathcal{G}_{t}\left( N,t_{0}\right) $; $d$ versus
time, averaged over 10 realizations}
\label{fig_1}
\end{figure}

\begin{figure}[tbp]
\begin{center}
\includegraphics {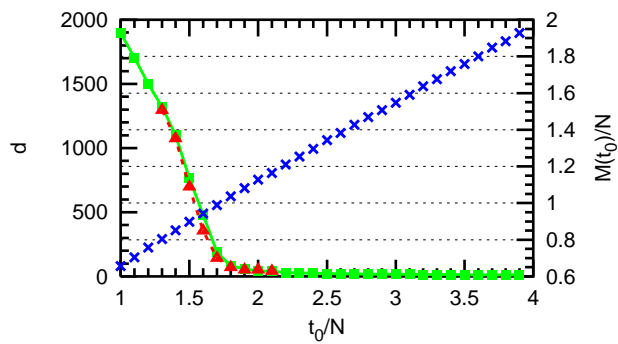} \vspace{-1cm}
\end{center}
\bf\caption{\rm Transition in the circle system; $N=10^{4}$, left axis: mean
diameter $d$ calculated at $t/N=10^{4}$ (boxes with lines) and $t/N=2\cdot
10^{4}$ (triangles with lines), right axis: effective non-base edges $%
M\left( t_{0}\right) $ (crosses) versus $t_{0}/N$ calculated at $t/N=10^{4}$%
; averaged over 10 realizations}
\label{fig_2}
\end{figure}

The first surprising observation is the collapse of the diameter from $%
const\cdot N$ to $const\cdot \log N$ for sufficiently large $t$, $t_{0}$ and
typical elements of $\mathcal{G}_{t}\left( N,t_{0}\right) $. This happens
despite the fact that only local edge formations where used. Fig.~\ref{fig_1}
shows the diameter versus $t$ for different $N$ and $t_{0}$. Our simulations
indicate that in the case $t_{0}/N\gtrsim 2$ one gets an asymptotic regime
in which the diameter is very small and shows no strong dependence on $N$.
Tab.~\ref{tab_1} shows for three choices of $t_{0}/N$ the mean diameter $%
\bar{d}$ in the asymptotic regime. The diameter $\bar{d}$ is approximately
proportional to $\log N$ which is one of the characteristics of small world
graphs. In the case $t_{0}/N=2$ an asymptotic regime with still small
diameter but large fluctuations starts approximately at $t/N\approx 10^{4}$.
Note that this value is much bigger than the collapse time in the cases $%
t/N=4,10$. It seems that $t_{0}/N$ $\approx 2$ is just the borderline where
a collapse of the diameter appears. This corresponds to the case where the
mean number of non-base edges equals $N$. For smaller values of $t_{0}$ we
found the diameter to stay of order $f\left( t_{0}\right) \cdot N$ with an
almost linear function $f$ (see Fig. \ref{fig_2}). This indicates an
interesting phase transition in the total number of edges reminiscent to the
famous  phase transition in the size of the largest component in classical
random graph spaces like the $G\left( N,M\right) $ -- the space of all
graphs with $N$ vertices and $M$ edges equipped with the uniform probability
measure.

\begin{table}[tbp]
\bf\caption{Mean diameter of the circle system}\rm
\label{tab_1}
\vspace{-0.5cm}
\begin{eqnarray*}
\begin{array}{l|l|l|l|l|l}
t_{0}/N & N & M(t_0)/N & \bar{d} & \bar{d}/\log _{10}N & \bar{d}_{random} \\ 
\hline
2 & 10^3 & 1.1  & 24.0 & 8.00 & 7.7\\
2 & 10^4 & 1.1  & 47.0 & 11.75 & 10.1\\
2 & 10^5 & 1.1  & 70.6 & 14.12 & 12.3\\ \hline
4 & 10^{3} & 2.0 & 8.3 & 2.78 & 6.1 \\ 
4 & 10^{4} & 2.0 & 11.6 & 2.90 & 7.7 \\ 
4 & 10^{5} & 2.0 & 15.0 & 3.00 & 9.2 \\ \hline
10 & 10^{3} & 4.4 & 5.1 & 1.70 & 4.2 \\ 
10 & 10^{4} & 4.4 & 6.7 & 1.68 & 5.7 \\ 
10 & 10^{5} & 4.4 & 8.2 & 1.64 & 7.0
\end{array}
\end{eqnarray*}
calculated by averaging the values $d(t)$ for $10^4 < t/N < 2\cdot 10^4$
($t_0/N = 2$) or $10^{3}<t/N<4\cdot 10^{3}$ ($t_0/N = 4, t_0/N = 10$) and
10 realizations; effective non-base edges $M(t_0)$;
the mean diameter $\bar{d}_{random}$ of $G_c (N,M(t_0))$ is calculated
by averaging over 100 realizations
\end{table}

\begin{figure}[tbp]
\begin{center}
\includegraphics {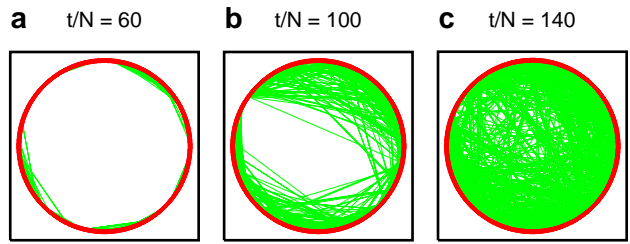} \vspace{-1cm}
\end{center}
\bf\caption{\rm Realization of a graph evolution using the circle
base edge system; $N=10^3$, $t_{0}/N=4$}
\label{fig_3}
\end{figure}

Fig.~\ref{fig_3} shows one realization of the time evolution of a typical
graph from $\mathcal{G}_{t}\left( N,t_{0}\right)$ with $t_{0}/N=4$ and $%
N=10^{3}$. It illustrates how the diameter of the graph becomes very small
only by using the IFF principle. Before the asymptotic state is reached an
interesting symmetry breaking can be observed: the appearance of a few
components with long range edges (taken the circle as a reference system)
and hence small diameter. The components itself are only connected via the
circle skeleton. It takes a relative long time till the components finally
merge and the circle is filled uniformly with edges. The same phenomenon can
be seen for larger values of $N$ and different $t_{0}/N$.

\begin{figure}[tbp]
\begin{center}
\includegraphics {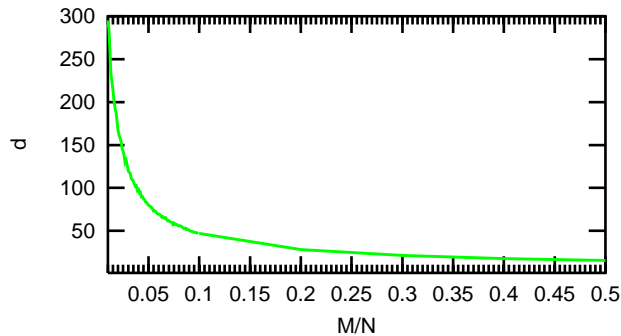} \vspace{-1cm}
\end{center}
\bf\caption{\rm Diameter $\bar{d}_{random}$ of $G_c(N,M)$ versus $M/N$;
averaged over 100 realizations}
\label{fig_4}
\end{figure}

It is interesting to compare our model with a version of the random graph
space $G\left( N,M\right)$ where an underlying circle is added to keep the
graph connected. To be precise let $G_{c}\left( N,M\right)$ be the random
graph space where $M$ edges are randomly added to a circle graph with $N$
vertices. Fig.~\ref{fig_4} shows the diameters $\bar{d}_{random}$ of the
average of a few samples of $G_{c}\left( N,M\right)$ as a function of $M/N$.
Here the diameter clearly decays like $\frac{const\cdot N}{M}$ and reaches
the $\log N$ regime already for very small values of $M/N$ ($\sim 0.05$).
Note that this value is much smaller than the threshold value $%
M/N\sim $ $0.5$ for the emergence of a giant component in $G\left( N,M\right)
$. Let us remark that the $G_{c}\left( N,M\right)$ model is very close in
spirit to the original Small World model of Watts and Strogatz. The main
difference is with respect to the clustering coefficient since the
underlying skeleton -- the circle -- has zero clustering coefficient.
Replacing the standard circle with next neighbor connections by one where
also next-next neighbor connections are edges would give an
essentially equivalent model to the Watts-Strogatz one.

\subsection{Fat tail of the degree distribution}

\begin{figure}[tbp]
\begin{center}
\includegraphics {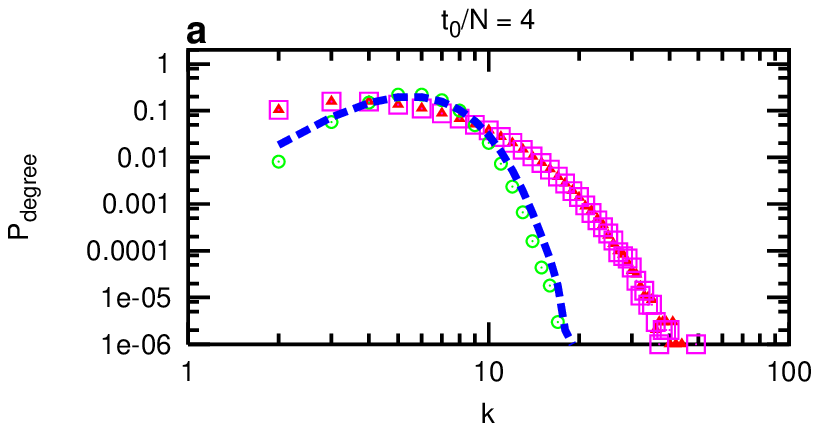} \vspace{-0.5cm}
\par
\includegraphics {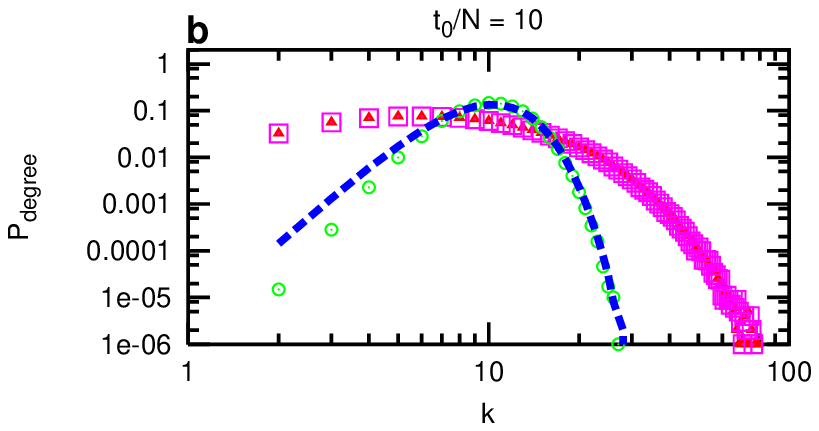} \vspace{-1cm}
\end{center}
\bf\caption{\rm Degree distribution using the circle system;
$N=10^5$, \textbf{a} $t_{0}/N=4$, $t/N=4$ (circles), $t/N=600$ (boxes),
$t/N=4000$ (triangles), 
\textbf{b} $t_{0}/N=10$, $t/N=10$ (circles), $t/N=600$ (boxes), $t/N=4000$
(triangles); the big dotted line shows the degree distribution of $%
G_{c}\left( N,M\left( t_{0}\right) \right) $, all values averaged over 10
realizations}
\label{fig_5}
\end{figure}

Another unexpected property of our model is the fat tail of the degree
distribution. Fig.~\ref{fig_5}a/b show the degree distribution for $N=10^{5}$
and $t_{0}/N=4\left( 10\right) $ at various times.

Let us compare the degree distribution in the asymptotic regime with those
of the random graph $G_{c}\left( N,M\left( t_{0}\right) \right) $, where $%
M\left( t_{0}\right) $ is the expectation of the number of edges in the IFF
random graph space $\mathcal{G}_{t}\left( N,t_{0}\right) $ (see Fig. \ref
{fig_2} for a plot of $M\left( t_{0}\right) /N$). The resulting degree
distributions are shown in Fig.~\ref{fig_5}a and Fig.~\ref
{fig_5}b by big dotted lines. At the beginning (that is for $t\sim t_{0}$) the
IFF-model has essential the degree distribution of $G_{c}\left( N,M\left(
t_{0}\right) \right) $ but in the asymptotic regime the distributions differ
drastically since the IFF-model has gotten a fat tail. At the moment the
limited numerical data still don't allow to check whether it is really a
power law.

\begin{figure}[tbp]
\begin{center}
\includegraphics {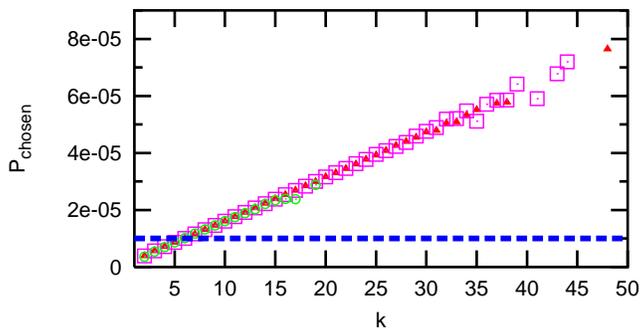} \vspace{-1cm}
\end{center}
\bf\caption{\rm Probability of a vertex to be chosen versus its degree;
$N=10^5$, $t_{0}/N=4$ and $t/N=4$ (circles), $t/N=600$ (boxes), $t/N=4000$
(triangles), the big dotted line shows the constant probability in $G\left(
N,M\right) $ averaged over 10 realizations; basic edges system: circle }
\label{fig_6}
\end{figure}

We want to close this section with a remark about the degree preferences if
a vertex $x$ was chosen in the formation of a new edge. Although never
explicitly included in the model there is a strong numerical evidence (see
Fig.~\ref{fig_6}) that the probability of a vertex to be chosen is
proportional to its degree. That is one of the basic assumptions in the
Albert--Barab\'asi model. A heuristic explanation for this property in our
model is the following argument based on two independence assumptions. Let $%
N_{1}$ and $N_{2}$ be the expected values for the first and second
neighborhood sizes. Assume that the conditional expectation of $N_{2}\left(
x\right) $ for $x$ having degree $k$ equals $k\cdot N_{1}$ and assume
further that for $z\in N_{2}\left( x\mid d\left( x\right) =k\right)$ the
expected value of $N_{2}\left( z\right)$ equals $N_{2}$. Clearly with these
conditions one can easily compute the probability of a vertex $x$ with
degree $k$ to be chosen within the process of an edge formation to be equal
to $\frac{k\cdot N_{1}}{N}\cdot \frac{1}{N_{2}}$ .

\subsection{Clustering properties}

\begin{figure}[tbp]
\begin{center}
\includegraphics {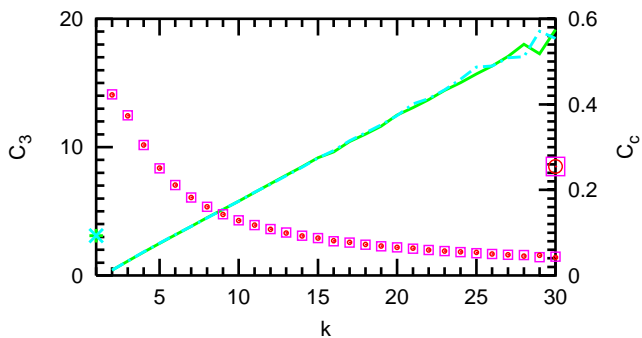} \vspace{-1cm}
\end{center}
\bf\caption{\rm Mean number of triangles $C_{3}$ and mean clustering
coefficient $C_{c}$ per vertex versus degree $k$, $N=10^5$, $t_{0}/N=4$,
the big symbol
shows the mean values of all vertices independent of their degree, all
values averaged over 10 realizations, the basic edges system forms a circle:
number of triangles (left axes): $t/N=600$ (dashed line/plus sign), $t/N=4000
$ (dotted line/cross); clustering coefficient (right axes): $t/N=600$
(boxes), $t/N=4000$ (circles) }
\label{fig_7}
\end{figure}

A basic quantity to measure the local clustering around a vertex $x$ is the
number of triangles $C_{3}\left( x\right) $ containing $x$ as a vertex. Note
that for classical random graph spaces the expectation of this number is is
of order $N^{-1}$. We are interested in the averaged number of triangles in
the asymptotic regime of the IFF-model. Fig.~\ref{fig_7} shows the mean
number of triangles $C_{3}$ for all vertices with a fixed degree for $t/N=600
$ and $t/N=4000$. The mean number of triangles per vertex (independent of its
degree) is shown by the big symbols on the figure frame. The plot for the two
time-values coincides practically. The number of triangles increases nearly
linear with the degree $k$. The reason for the fluctuations for high $k$ is
the low number of vertices with such high degrees. Fig.~\ref{fig_7} shows
also the mean clustering coefficient $C_{c}$ which is directly connected to
the mean number of triangles. Namely the clustering coefficient $C_{c}\left(
x\right) $ of a vertex $x$ with degree $k$ is given by the normalized
triangle coefficient $\frac{2C_{3}\left( x\right) }{k\left( k-1\right) }$ .

\begin{table}[tbp]
\bf\caption{Mean value of $\varphi^{(n)}$ of the circle system}\rm
\label{tab_2}
\vspace{-0.5cm}
\begin{eqnarray*}
\begin{array}{c|r|r|r}
t_0/N & t/N & \varphi^{(2)} & \varphi^{(3)} \\\hline
4  & 600  & 0.28 & 0.15\\
4  & 4000 & 0.29 & 0.15\\ \hline
10 & 600  & 0.30 & 0.16\\
10 & 4000 & 0.30 & 0.16\\ \hline
\end{array}
\end{eqnarray*}
Calculated with $N=10^5$ and after $t$ steps, averaged over 10 realizations
\end{table}

There is another interesting quantity, defined via the second shortest
distance between two vertices, which characterizes additional clustering
properties. For pairs of vertices with a common edge let $d_{2}\left(
x,y\right) $ be the distance between $x$ and $y$ after removal of the
connecting edge. Define for a given graph $G$ the quantity $\varphi^{(n)}
\left(G\right)$ as the fraction of pairs of vertices with distance $1$ whose
$d_{2}$-distance is larger $n$ (for a given random graph space $\mathcal{G}$
let $\varphi^{(n)} \left( \mathcal{G}\right)$ the expectation of
$\varphi^{(n)}$). In
case when the diameter of a random graph space is small due to the presence
of independently generated edges the $\varphi^{(2)}$ value is usually
very large since no short second-shortest paths between two vertices
connected by an
independently generated edge exist. This remains also true for the
Watts-Strogatz Small World graph. In our model we get by the very
construction process a small value of $\varphi$ (3) similar to the situation
met in real networks (see Tab. 2).

\subsection{Replacing the circle by a torus}

Up to now, the system of base edges forms a circle. It is natural to ask if
one gets qualitatively the same results with a two-dimensional system of
base edges like a torus lattice. This means that at the beginning the
vertices are connected in form of a torus or a lattice with periodic
boundaries. The algorithm to generate (and define) $\mathcal{G}_{t}\left(
N,t_{0}\right)$ remains the same.

For the diameter one gets again an asymptotic regime with collapse up to
logarithmic size but the asymptotics starts earlier as in the
``circle''-case. Even for $t_{0}/N=2$ we get an asymptotic regime starting
at $t/N\approx 200$. Tab.~\ref{tab_3} shows the mean diameter $\bar{d}$ in
the case of a torus base for various values of $t_{0}$ and $N$.

\begin{table}[tbp]
\bf\caption{Mean diameter of the torus system}\rm
\label{tab_3}
\vspace{-0.5cm}
\begin{eqnarray*}
\begin{array}{c|r|r|r|r|r}
t_0/N & N & M(t_0)/N & \bar{d} & \bar{d}/\log_{10}N & \bar{d}_{random} \\ 
\hline
2 & 30^2 & 1.5 & 7.9 & 2.67 & 5.6 \\ 
2 & 100^2 & 1.5 & 14.1 & 3.53 & 7.0 \\ 
2 & 300^2 & 1.5 & 27.6 & 5.57 & 8.4 \\ \hline
4 & 30^2 & 2.8 & 5.1 & 1.73 & 4.7 \\ 
4 & 100^2 & 2.8 & 6.7 & 1.68 & 6.0 \\ 
4 & 300^2 & 2.8 & 8.0 & 1.61 & 7.0 \\ \hline
10 & 30^2 & 6.6 & 4.0 & 1.35 & 4.0 \\ 
10 & 100^2 & 6.7 & 5.0 & 1.25 & 4.8 \\ 
10 & 300^2 & 6.7 & 6.0 & 1.21 & 5.5
\end{array}
\end{eqnarray*}
Calculated with the values of the mean diameters in the range $1000 < t/N <
4000$ averaged over 10 realizations; the value $\bar{d}_{random}$ is
calculated by averaging over 100 realizations
\end{table}

\begin{figure}[tbp]
\begin{center}
\includegraphics {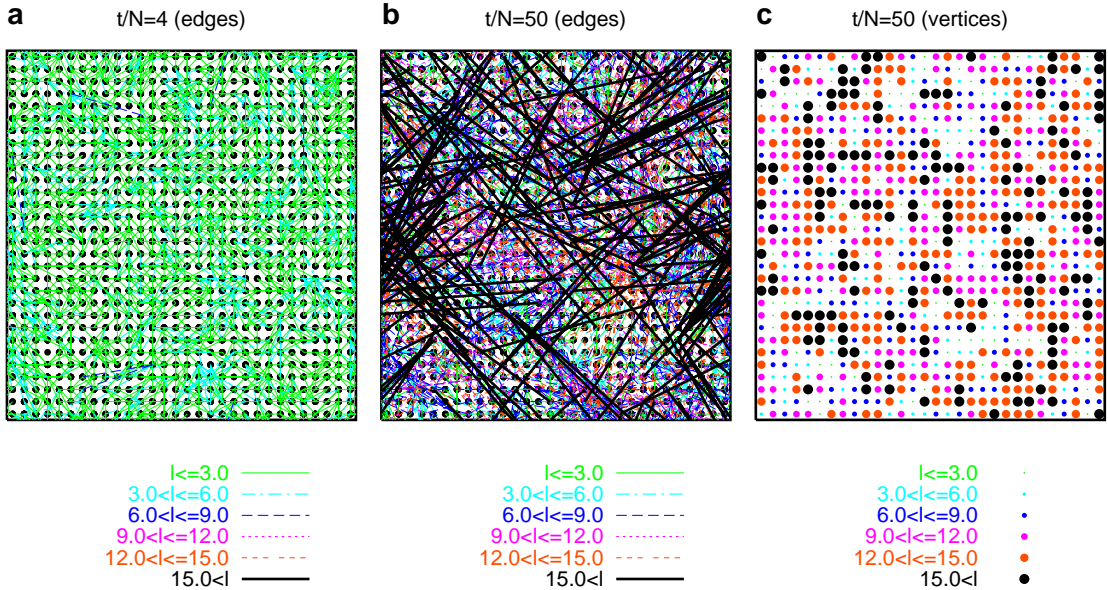} \vspace{-1cm}
\end{center}
\bf\caption{\rm Realization using the torus system; $N=30^{2}$, $t_{0}/N=4$; 
\textbf{a,b} the edges at different times are plotted, the dots characterize
the vertices, the color of an edge characterizes its length $l$ ; \textbf{c}
the dots characterize the vertices, the color characterize the length $l$ of
the longest edge of the vertex}
\label{fig_8}
\end{figure}

There is another difference to the one-dimensional case which can be seen in
the realization shown in Fig.~\ref{fig_8}: the graph evolution is more
uniformly with a two-dimensional base edge system than with a
one-dimensional (compare to Fig.~\ref{fig_3}).

We obtain qualitatively the same results as in the one-dimensional case with
respect to the degree distribution and the clustering properties for $%
N=300^{2}$ and $t_{0}/N=4$, $t_{0}/N=10$. In the asymptotic regime one sees
again a clear fat tail distribution. The distribution of the number of
triangles and the clustering coefficient for $N=300^{2}$ and $t_{0}/N=4$
gives qualitatively the same results as in the ``circle''-case, i.e. the
distributions are nearly equal for $t/N=600$ and $t/N=4000$ and we have a
linear dependence of the number of triangles on the degree $k.$

\subsection{Removal of the base edge system}

\begin{figure}[tbp]
\begin{center}
\includegraphics {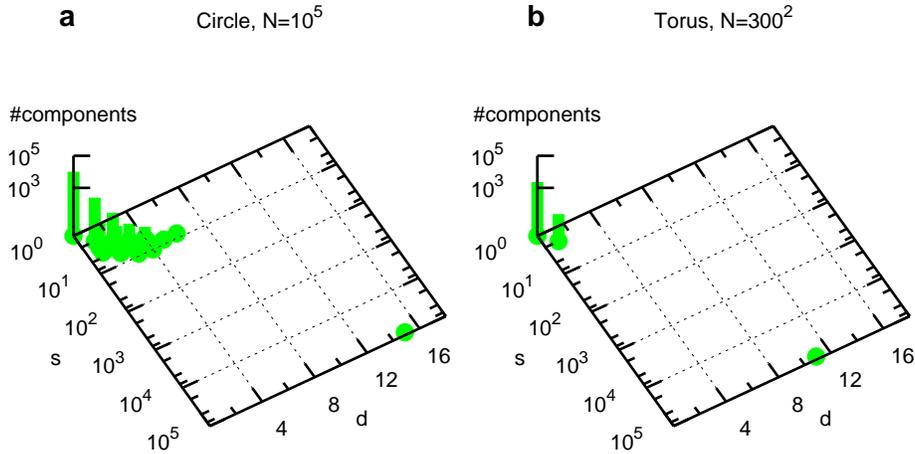} \vspace{-1cm}
\end{center}
\bf\caption{\rm Number of components with a given number of vertices
and diameter,
after deleting the basic edges, $t/N=1000$, $t_{0}/N=4$, one realization }
\label{fig_9}
\end{figure}

\begin{figure}[tbp]
\begin{center}
\includegraphics {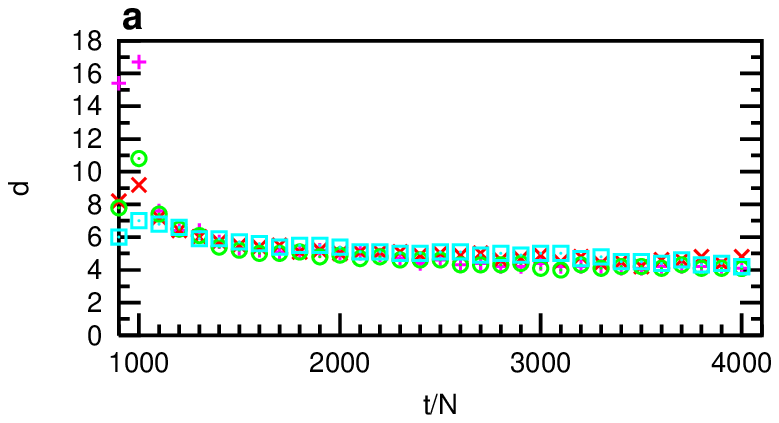}
\par
\includegraphics {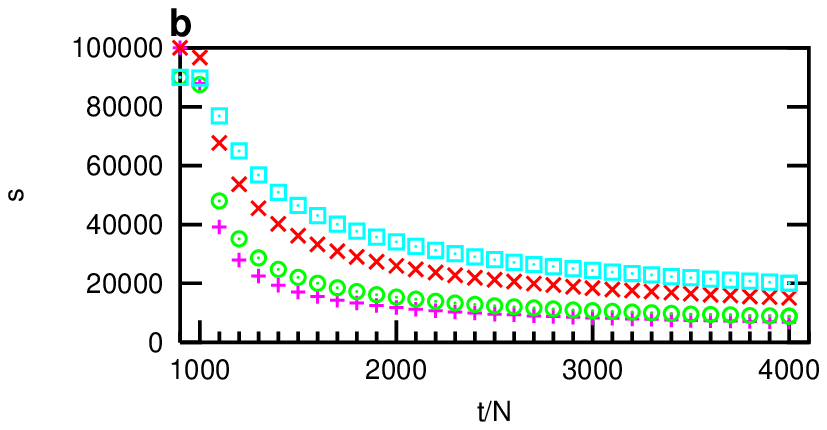} \vspace{-1cm}
\end{center}
\bf\caption{\rm Parameters of the biggest component after deleting the basic
edges; \textbf{a} time dependence of its diameter $d$; \textbf{b} time
dependence of its size $s$; circle, $N=10^5$, $t_{0}/N=4$ (plus signs);
circle, $N=10^5$, $t_{0}/N=10$ (crosses); torus, $N=300^{2}$, $t_{0}/N=4$
(circles); torus, $N=300^{2}$, $t_{0}/N=10$ (boxes); the symbols on the left
vertical axes indicate the values of the parameters directly before deleting
the basic edges; all values averaged over 10 realizations}
\label{fig_10}
\end{figure}

Finally we want to discuss the stability of the graphs in the asymptotic
regime under removal of the circle edges or torus edges, i.e. at $t/N=1000$
we remove the base edges and examine in which way the graph breaks into
connected components and how these components evolve by continuing using our
algorithm. Fig.~\ref{fig_9} shows the number of connected components with
different size $s$ (number of vertices) and different diameter $d$ for $%
t_{0}/N=4$ directly after deleting the base edges at $t/N=1000$ for the
``circle'' and the ``torus'' case. There is still one big component. The
time evolution of the biggest component can be seen in Fig.~\ref{fig_10}
which shows its size $s$ and diameter $d$ versus time. An interesting result
is that the diameter of the biggest component is nearly constant for $%
1500<t/N<4000$, nearly independent of $t_{0}/N$ and also nearly independent
of using a circle or a torus as base edge system. Let us note that without
the skeleton there can be no merging of different components by the very
nature of the IFF principle. Therefore, as can easily be seen, the only
stable asymptotic configurations are components which are totally connected
(i.e. every vertex has an edge to all other vertices in the component) since
no further changes in such components can happen. But it is very likely that
the time scale till this phenomenon can be seen is huge compared to the time
scales we where studying.

To compare with the classical random graph situation observe that the
removal of the circle in $G_{c}\left( N,M\right) $ gives just the model $%
G\left( N,M\right) .$ The component distribution for this random graph space
is well known. Especially for $M>N/2$ there is always a giant component of
size $const\cdot N.$ But in contrast to our model, the diameter $d$ of the
biggest component depends via the constant on $M\left( t_{0}\right) $.

%
%

\section{Summary and outlook}

We have presented a model where the iteration of the ''My Friends are Your
Friends (FF)'' principle has produced random graphs with small world
properties (logarithmic diameter, high clustering) and a fat tail
distribution for the degree. The model used a fixed regular graph of large
diameter as a skeleton, reminiscent to a pre-given geographical structure
with pre-given neighborhood relations for vertex pairs. But in contrast to
the small world graphs by Watts and Strogatz no random, global edge
formation processes are involved to obtain small diameter. The high
clustering is an immediate consequence of the FF -principle. The fat tail
distribution is surprising since no preferential choice mechanisms are
contained in the graph generation algorithm. So far our investigations are
entirely numerical and clearly a more theoretical explanation of the
observed phenomena is the desired next step.

There are several natural variants of the model which we would like to
mention shortly.

First, in our model the total number of edges is kept fixed after the build
up phase. Instead of that one could use probabilistic rules which keep the
number of edges only fixed in mean. This would match better real situations
where the FF-principle is of relevance.

Second, like in the Albert \& Barab\'asi network \cite{3}, a growing number
of vertices could be considered. Growth should happen here in form of
offsprings of already existing vertices to be able to apply the FF-principle
to the ''newcomers''. In this situation it could well be the case that even
without the iteration of the FF-principle -- every vertex when entering the
network forms just once a number of edges according to the FF-rule -- Small
World graphs plus fat tail for the degree distribution are obtainable. This
model variant is actually the typical situation for the growth and formation
of the network of citations or collaboration.


\begin{thebibliography}{9}
\bibitem{1}  D.J. Watts, S.H. Strogatz : Nature \textbf{393},\textbf{\ }440
(1998)

\bibitem{5}  D.J. Watts:\emph{\ Small Worlds: The dynamics of networks
between order and randomness}, Princeton University Press, 1999

\bibitem{4}  R. Albert , A.-L. Barab\'asi : \emph{Statistical Mechanics of
Complex Networks}, Reviews of Modern Physics, \textbf{74},\textbf{\ }47
(2002), arXiv:cond-mat/0106096

\bibitem{3}  A.-L. Barab\'asi, R. Albert : Science, \textbf{286},\textbf{\ }%
509 (1999)

\bibitem{2}  Ph. Blanchard, T. Krueger : \emph{The ''Cameo Principle'' and the
origin of scale free graphs in social networks }, (2003),
arXiv:cond-mat/0302611
\end{thebibliography}
\end{document}